\documentclass[11pt]{article}
\usepackage{graphicx,verbatim,array,multicol,palatino,amssymb,amsfonts, amsmath, subfigure, epsfig}
\usepackage{color}
\usepackage{chicago}
\usepackage{url}
\usepackage{bbm}

\def\jump{\vskip3mm\noindent}

\newtheorem{theo}{Theorem}[section]
\newtheorem{cor}{Corollary}
\newtheorem{assump}{Assumption}

\begin{document}

\null\vskip2cm
\begin{center}

{\LARGE
\bf Simultaneous adjustment of bias and coverage probabilities for confidence intervals}
\vskip5mm
\begin{large}

{\sc P. Men\'endez}
\footnote{
School of Mathematics and Physics,
University of Queensland, St Lucia, 4072, AUSTRALIA}
{\sc Y. Fan},
\footnote{
School of Mathematics and Statistics,
University of New South Wales, Sydney, 2052, AUSTRALIA}
{\sc P. H. Garthwaite}
\footnote{Department of Mathematics and Statistics, Open University, Milton Keynes, MK7 6AA, U.K.}
and
{\sc S. A. Sisson }$^2$\\

\end{large}
\vskip5mm

\today\vskip2mm

{\sc Abstract}\vskip2mm
\end{center}

\noindent A new method is proposed for the correction of confidence intervals when the original interval does not have the correct nominal coverage probabilities in the frequentist sense. The proposed method is general and does not require any distributional assumptions. It can be applied to both frequentist and Bayesian inference where interval estimates are desired. We provide theoretical results for the consistency of the proposed estimator, and give two complex examples, on confidence interval correction for composite likelihood estimators and in approximate Bayesian computation (ABC), to demonstrate the wide applicability of the new method. Comparison is made with the double-bootstrap and other methods of improving confidence interval coverage.

\jump
\vskip4mm

\section{Introduction}%

Interval estimates are typically intended to have a specified level of coverage. This is true, for example, of both frequentist confidence intervals and Bayesian credible intervals. However, for many problems the coverage of a confidence or credible interval will only equal its nominal value asymptotically, and coverage can be poor even for quite large samples in some situations.
In many complex problems, there can be inherent bias which can be difficult to quantify or calculate. This can arise, for example, in composite likelihood problems \shortcite{varin2011overview} and approximate Bayesian computation  \shortcite{sisson+f11}.
In this paper we propose a novel procedure for adjusting interval estimates that has wide application and will typically reduce the bias in their coverage.
\\

\noindent The procedure assumes that the mechanism that generated the sample data could be simulated if population parameters were known. These parameters are estimated by sample statistics derived from real data, and then pseudo-samples are drawn from the estimated population distribution. From each pseudo-sample a confidence/credible interval is determined for the quantity of interest. The frequentist bias in these intervals is calculated and then used to adjust the interval estimate given by the real data.
\\

\noindent The method has similarities to the double-bootstrap \cite{davison1997bootstrap}, in which a bias correction is applied to a bootstrap interval by re-sampling from a bootstrap distribution. A difference in our method is that it involves only one level of sample generation, which makes it computationally less demanding, although the saving may dissipate if computationally demanding methods (such as Markov chain Monte Carlo) are used to obtain the interval estimates from sample data.\\

\noindent
Several authors have  looked at the problem of computing confidence intervals with the correct coverage properties. Much of this work is based on variations of bootstrap procedures.
In particular,
\citeN{hall1986bootstrap} considered the bootstrap in terms of Edgeworth expansions, and  \citeN{beran1987prepivoting}  provided a method for approximate confidence sets, by using the bootstrap (or asymptotic theory) to estimate the relevant quantiles. This so-called {\it pre-pivoting}, based on an estimated bootstrap cumulative distribution function, is iterated to produce improved coverage.
\citeN{efron1987better} introduced a method to correct the coverage of a bootstrap confidence interval within the bootstrap itself, and  \citeN{martin1990bootstrap} proposed an iterated bootstrap procedure to obtain better bootstrap coverage.
The so called double bootstrap, where one or more additional levels  of bootstrap are conducted to adjust confidence limits, was disccussed by \shortciteN{diciccio1992fast}.  In a non-bootstrap procedure, \citeN{garthwaite1992generating} proposed a method to compute confidence intervals based on Monte Carlo simulations of the Robbins-Monro search process.  In the context of autoregressive models, \citeN{hansen1999grid} presented a method based on bootstrap replications over a grid to compute confidence intervals in situations where standard bootstrap methods fail. There are also several approaches for computing confidence intervals in the presence of nuisance parameters \cite{kabaila1993some,kabaila2000computable,Lloyd2011}.
\\

\noindent In some of our examples, we apply our procedure to reduce bias in the coverage of Bayesian credible intervals.
This may not seem intuitive, since {\em coverage} is a frequentist property while a Bayesian interval may reflect personal probabilities.
However, there are many situations where posterior distributions should preferably be well calibrated. These include inference with objective or probability matching prior distributions, the verification of Bayesian simulation software \shortcite{cook+gr06} and techniques and diagnostics in likelihood-free Bayesian inference \shortcite{fearnhead+p12,prangle+bps12}.
\\

\noindent In Section \ref{secCov}  we describe the proposed method and give theoretical results related to it, illustrating them through
simulated examples and comparing the resulting intervals with bootstrap intervals.
In Section \ref{examples} we apply our method to two more complex, real analyses. One of these involves estimation with composite likelihoods, which is known to produce confidence intervals that are too narrow, and the other involves approximate Bayesian computation, which typically gives larger posterior credibility intervals than desired. Some concluding comments are given in Section \ref{secConc}.

\section{Coverage correction for confidence intervals}%
\label{secCov}

Suppose we are interested in estimating an equal-tailed $100(1-\alpha)\%$ confidence interval for some parameter $\theta \in \Theta \subseteq {\mathcal R}$. Thus for observed data ${\bf x}$, we seek an estimate $L_{c}({\bf x})$, such that
$$P(\theta \leq L_{c}({\bf x}) ) = \alpha/2,$$
where $L_c({\bf x})$ denotes the lower limit of the interval. Similarly for the upper limit, we seek an estimate $U_{c}({\bf x})$, such that,
$$P(\theta \geq U_{c}({\bf x}) ) = \alpha/2.$$

\noindent In the frequentist setting, the parameter $\theta$ is considered a fixed quantity and the expressions above  are written in terms of pivotal functions of the data, ${\bf x}$. In the Bayesian setting, the credible interval is computed from the quantiles of the posterior distribution of $\theta$.
In an abuse of notation, we will use the above notations
in both frequentist and Bayesian cases. \\

\noindent
Suppose that we have a method of obtaining estimates, $L({\bf x})$ and $U({\bf x})$, of the correct lower and upper interval bounds, $L_c({\bf x})$ and $U_c({\bf x})$.
We do not assume that these estimates produce the correct coverage probability.
However, we do assume that the population parameters, $\theta$, can be well approximated from the data.
Our goal is to provide a method that gives adjustments to $L({\bf x})$ and $U({\bf x})$ that improve the interval's coverage.
We first give theoretical results for the proposed methodology, and then give details of its implementation.

\subsection{Theoretical results}%
\label{section:theo}

\begin{assump}\label{ass1}
We suppose that the observed data ${\bf x}$ come from the model given by $f({\bf x}| \theta)$, $\theta\in \Theta$.
For any $\theta \in \Theta$, we assume that it is possible to simulate from $f(\cdot|\theta)$.
\end{assump}

\begin{assump} \label{ass2}
Given $\theta$ and data ${\bf x} \sim f({\bf x}|\theta)$ there exists a consistent estimator $\tilde{\theta}$ of $\theta$.
\end{assump}

\noindent Assumption \ref{ass1} requires that we are able to simulate replicate data from the model given the values of the parameters.
Assumption \ref{ass2}  requires that we have a good estimator for $\theta$, so that interval estimates obtained using $\tilde{\theta}$ converge to those estimates obtained using the population parameter $\theta$, as the amount of data gets large.
\\

\noindent In the following, we only require the lengths of the intervals to be consistent. Consequently,  Assumption 2 is not always necessary. For example, this occurs if $\theta$ represents a location parameter whose
confidence interval has a length that is independent of $\theta$ (see later example).  In the frequentist setting, the maximum likelihood estimator of $\theta$ is  consistent and  unbiased in many finite sample situations.
In the Bayesian setting, the posterior distribution is consistent under mild assumptions, and the posterior mean estimate of $\theta$ is asymptotically unbiased. However, in both cases, finite sample bias in  $\tilde{\theta}$ may render our method less accurate.
\\

\begin{theo}\label{thm1}
For some $\theta$ and ${\bf x}\sim f({\bf x}| \theta)$, let $L({\bf x})$ be an estimator of the lower limit of a $100(1-\alpha)\%$ level confidence interval, and suppose that
$$P \{ \theta < L ( {\bf x}) \} \neq \alpha/2.$$
Let $G_{\{W\}}$ denote the distribution function of a random variable $W$.
Consider the new estimator
\begin{equation}\label{eqnLT}
L_c({\bf x} ) = L({\bf x}) + \xi_{\alpha/2},
\end{equation}
where $\xi_{\alpha/2}$ is the $\alpha/2$-th  quantile
of the distribution function $G_{ \{ \theta-L({\bf x}) \} }$, so that  $G_{ \{ \theta-L({\bf x}) \} } (\xi_{\alpha/2})  =\alpha/2$.
Then the new estimator, $L_c({\bf x} )$, will have the correct coverage probability
$$P \{ \theta < L_c({\bf x}) \}  = \alpha/2.$$

\end{theo}

\noindent{\bf Proof:} See Appendix.
\\

\noindent From the above theorem, it can then be seen that for the estimator of the upper limit of a $100(1-\alpha)\%$ confidence interval, $U( {\bf x})$, we can write
\begin{equation}\label{eqnUT}
U_c({\bf x})=U( {\bf x}) + \xi_{1-\alpha/2},
\end{equation}
where $\xi_{1-\alpha/2}$ is the $(1-\alpha/2)$-th  quantile of the distribution function  $G_{ \{ \theta-U({\bf x}) \} }$, so that $G_{ \{ \theta-U({\bf x}) \} } (\xi_{1-\alpha/2})=1-\alpha/2$. In this case, we then have that
$$P \{ \theta >  U_c({\bf x}) \}  = \alpha/2.$$
\\

\begin{theo}\label{thm2}
For some $\theta$ and observed data ${\bf x}\sim f({\bf x}| \theta)$, suppose the lower limit of a $100(1-\alpha)\%$ confidence interval $L( {\bf x})$ can be obtained, and that this estimate does not necessarily give the correct coverage probability.
Suppose that $\tilde{\theta}\in \mathcal{R}$ is a consistent estimator of $\theta$, evaluated using the data ${\bf x}$.
Let ${\bf y}_1,\ldots, {\bf y}_n$ be $n$ replicate datasets simulated independently from $f(\cdot |\tilde{\theta})$,
and denote the corresponding lower confidence limits by
$L_1({\bf y_1}),\ldots,L_n({\bf y_n})$, obtained in the same manner as $L({\bf x})$.
Define
 $$\hat{G}_{  \{ \tilde{\theta}-L({\bf y})\}} (\epsilon) = \frac{1}{n}\sum_{i=1}^n I_{\{\tilde{\theta}-L_i({\bf y_i}) < \epsilon\}}$$
as the empirical distribution of $\tilde{\theta}-L({\bf y})$ based on the observed values of $\tilde{\theta}-L_i({\bf y_i}), i=1,\ldots,n$.
If we define
\begin{equation}\label{eqnL}
\tilde{L}_c({\bf x} ) = L( {\bf x}) + \hat{\xi}_{\alpha/2}
\end{equation}
where $\hat{\xi}_{\alpha/2}=\hat{G}^{-1}_{ \{\tilde{\theta}-L({\bf y})\}}(\alpha/2)$, then  $\tilde{L}_c({\bf x} ) $ is a consistent estimator of $L_c({\bf x} )$, as  defined in
Equation (\ref{eqnLT}).
\end{theo}
\noindent{\bf Proof:} See Appendix.
\\

\noindent In combination, Theorems \ref{thm1} and \ref{thm2} state that if we simulate data ${\bf y_1},\ldots,{\bf y_n}\sim f({\bf y} |\tilde{\theta})$ and subsequently
obtain the confidence limits $L_1({\bf y_1}),\ldots,L_n({\bf y_n})$ in the same way as for the original data ${\bf x}$,
then we can correct the bias in the original lower limit estimate, $L({\bf x})$, by addition of  the $\alpha/2$-th sample quantile of $\tilde{\theta}-L_1({\bf y_1}),\ldots,\tilde{\theta}-L_n({\bf y_n})$.

\begin{cor}\label{cor1}
Under the assumptions in Theorem \ref{thm2}, a central limit theorem holds for $\tilde{L}_c({\bf x})$. Specifically, for all $\alpha \in (0,1)$, $\theta\in {\mathcal R}$ and ${\bf x} \sim f({\bf x}|\theta)$, we have that
$$\sqrt{n}(\tilde{L}_c\big( {\bf x})-L_c({\bf x})\big)G_{ \{ \theta-L({\bf x}) \} }^{'}(\xi_{\alpha})\longrightarrow N(0,\alpha(1-\alpha))$$
as $n \rightarrow \infty$,
where
$G_{ \{ \theta-L({\bf x}) \} }^{'}(\xi)=\frac{\partial}{\partial \xi} G_{ \{ \theta-L({\bf x}) \} }(\xi)$, and
$\xi_{\alpha}$  is the $\alpha$-th quantile of $G_{ \{ \theta-L({\bf x}) \} } $.
\end{cor}
\noindent{\bf Proof:} The result follows immediately from Equation (\ref{eqnCLT}) of the proof for Theorem \ref{thm2} (see Appendix).
\\

\noindent The above theoretical results provide a simple way of estimating corrections to the lower and upper confidence limits that will produce the correct nominal coverage
probability. In addition,
these estimators are consistent and asymptotically normal.

\subsection{Correction procedure}%
\label{secCor}
In summary, the correction algorithm has the following steps:

{\it
\begin{description}
\item[Step 1] Obtain $L( {\bf x})$ and $U({\bf x})$, the upper and lower limits of the desired $100(1-\alpha)\%$ confidence interval for the parameter $\theta$, for an observed dataset ${\bf x}$.
\item[Step 2] Evaluate $\tilde{\theta}$ and generate $n$ independent datasets ${\bf y}_1,\ldots, {\bf y}_n\sim f({\bf y}| \tilde{\theta})$ from the model.
\item[Step 3] For each dataset ${\bf y}_i$, compute the  $100(1-\alpha)\%$ lower  and upper confidence limits,
$L_i({\bf y}_i)$ and $U_i({\bf y}_i)$,
for the parameter $\tilde{\theta}$, using the same method as in Step 1.
\item[Step 4] Set the corrected lower and upper limits to
\begin{eqnarray*}
	\tilde{L}_c({\bf x}) & = & L( {\bf x}) +   \hat{G}^{-1}_{ \{ \tilde{\theta}-L({\bf y})\} } (\alpha/2)\\
	\tilde{U}_c({\bf x}) & = & U( {\bf x}) + \hat{G}^{-1}_{ \{ \tilde{\theta}-U({\bf y})\} } (1-\alpha/2)
\end{eqnarray*}
where $\hat{G}^{-1}_{\{ W\}}(\alpha)$ denotes the $\alpha$-th sample quantile of the random variable $W$.
\end{description}
}

\vskip 0.3cm

\subsection{Simple examples}
\label{sec:simple}

We illustrate the above procedure with two simple examples.
In the first, we consider confidence interval correction for the mean parameter of a normal distribution with known variance. In the second example, it is assumed that the mean is known and that we are interested in the variance parameter.
\vskip 5mm

\noindent{\bf\it Example 1: Normal distribution with known variance}\\

\noindent Suppose that $\theta$ is the location parameter of a Normal distribution with unit variance, so that  ${x}_i \sim N(\theta, 1)$ where ${\bf x}=({x}_1,\ldots,{x}_m)$.  In this case, the maximum likelihood estimator is $\tilde{\theta}=\bar{x}=\sum_i {x}_i/m$.
For illustration, we suppose that the confidence interval we obtain for $\theta$ does not have the correct coverage, in that we obtain the equivalent confidence interval when data are generated from ${x}_i \sim N(\theta, (1+\epsilon)^2)$ with $\epsilon \geq 0$. The value of  $\epsilon$ controls the amount of error in the coverage probability.
Following the usual frequentist approach,
the $100(1-\alpha)\%$ confidence interval for $\theta$
is given by $L( {\bf x} )= \bar{{x}}-z_{\alpha/2}(1+\epsilon)/\sqrt{m}$ and $U( {\bf x})= \bar{{x}}+z_{1-\alpha/2}(1+\epsilon)/\sqrt{m}$, where $z_{\alpha}$ is the $\alpha$-th quantile of the standard normal distribution.
Clearly the correction for the interval when $\epsilon >0 $ is
$L_c( {\bf x} ) = L({\bf x} ) +z_{\alpha/2}\epsilon/\sqrt{m}$ and
$U_c({\bf x} ) = U({\bf x} ) - z_{1-\alpha/2}\epsilon/\sqrt{m}$.
\\

\begin{figure}
\includegraphics[scale=0.8]{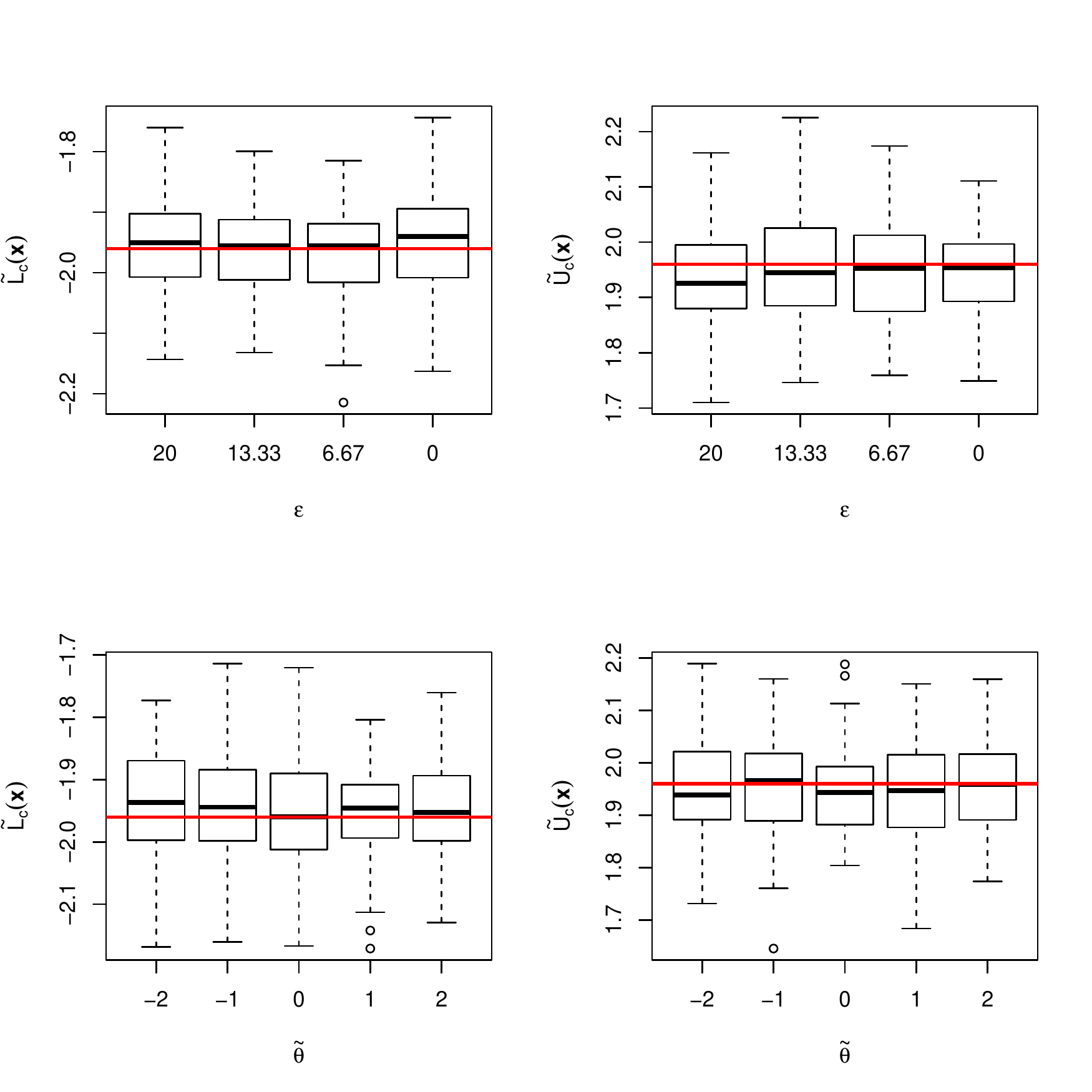}
\caption{\small Lower $\tilde{L}_c({\bf x})$ and upper $\tilde{U}_c({\bf x})$ corrected confidence limit estimates for 100 replicated analyses for the normal location model.
Top plots show the corrected limits
for  $\epsilon=20, 13.33, 6.67, 0$ with $\tilde{\theta}=\bar{{x}}$. Bottom plots show the corrected limits for $\tilde{\theta}=-2,-1,0,1, 2$ with  $\epsilon=1$. The horizontal lines represent the $0.025$-th (left plots) and $0.975$-th (right plots) percentiles of a standard normal distribution.}
\label{FigNOR}
\end{figure}

\noindent Figure~\ref{FigNOR} displays the results of the correction procedure for a 95\% confidence interval based on 100 replicate analyses. Each analysis is based on samples of size $m=20$ with $\theta=0$, so that ${x}_1,\ldots,{x}_m\sim N(0,1)$, and $n=100$ replicated samples ${\bf y}_1,\ldots,{\bf y}_n$ with elements drawn from
$N(\tilde{\theta},1)$.
Figure~\ref{FigNOR} (top plots) illustrates the corrected confidence limits $\tilde{L}_c(\bf x)$ and $\tilde{U}_c(\bf x)$ for a range of error term values, $\epsilon$. Clearly the correction produces an unbiased adjustment, as the boxplots are centred on the true confidence bounds (the horizontal line) in each case. Further, the performance of the method produces qualitatively the same corrected interval limits, irrespective of the value of $\epsilon$.
\\

\noindent The bottom plots display the corrections $\tilde{L}_c(\bf x)$ and $\tilde{U}_c(\bf x)$ with $\epsilon=1$ fixed, for a range of values of $\tilde{\theta}$.
For this example, choosing $\tilde{\theta}$ to be any arbitrary value will result in the same quality of unbiased correction. This arises as the distributions of $\tilde{\theta}-L({\bf y})$ and $\tilde{\theta}-U({\bf y})$ do not change with $\tilde{\theta}$, so that the confidence intervals all have the same width as $\tilde{\theta}$ varies. As this is a location parameter only analysis, this is one case where Assumption 2 is not required to produce a consistent adjustment (see Section \ref{section:theo}).
\\

\noindent{\bf\it Example 2: Normal distribution with known mean}\\%

\noindent Suppose now that $\theta$ is the scale (variance) parameter of a Normal distribution with mean zero, so that ${x}_i\sim N(0,\theta)$.
Here we specify  $\tilde{\theta}=S^2=\frac{1}{m-1}\sum_{i=1}^m ({x}_i - \bar{{x}})^2$ as the sample variance.
In this setting, suppose that the regular confidence limits for $\theta$ are biased downwards by a constant value $\epsilon>0$.
Specifically, the $100(1-\alpha)\%$ confidence interval for $\theta$ is given by $L({\bf x})=\frac{(m-1)S^2}{\chi^2_{1-\alpha/2;m-1}}-\epsilon$ and  $U({\bf x})=\frac{(m-1)S^2}{\chi^2_{\alpha/2;m-1}}-\epsilon$,  where
$\chi^2_{\alpha,k}$ denotes the $\alpha$-th percentile of a $\chi^2_k$ distribution with $k$ degrees of freedom.
\\

\begin{figure}
\begin{center}
\includegraphics[scale=0.6]{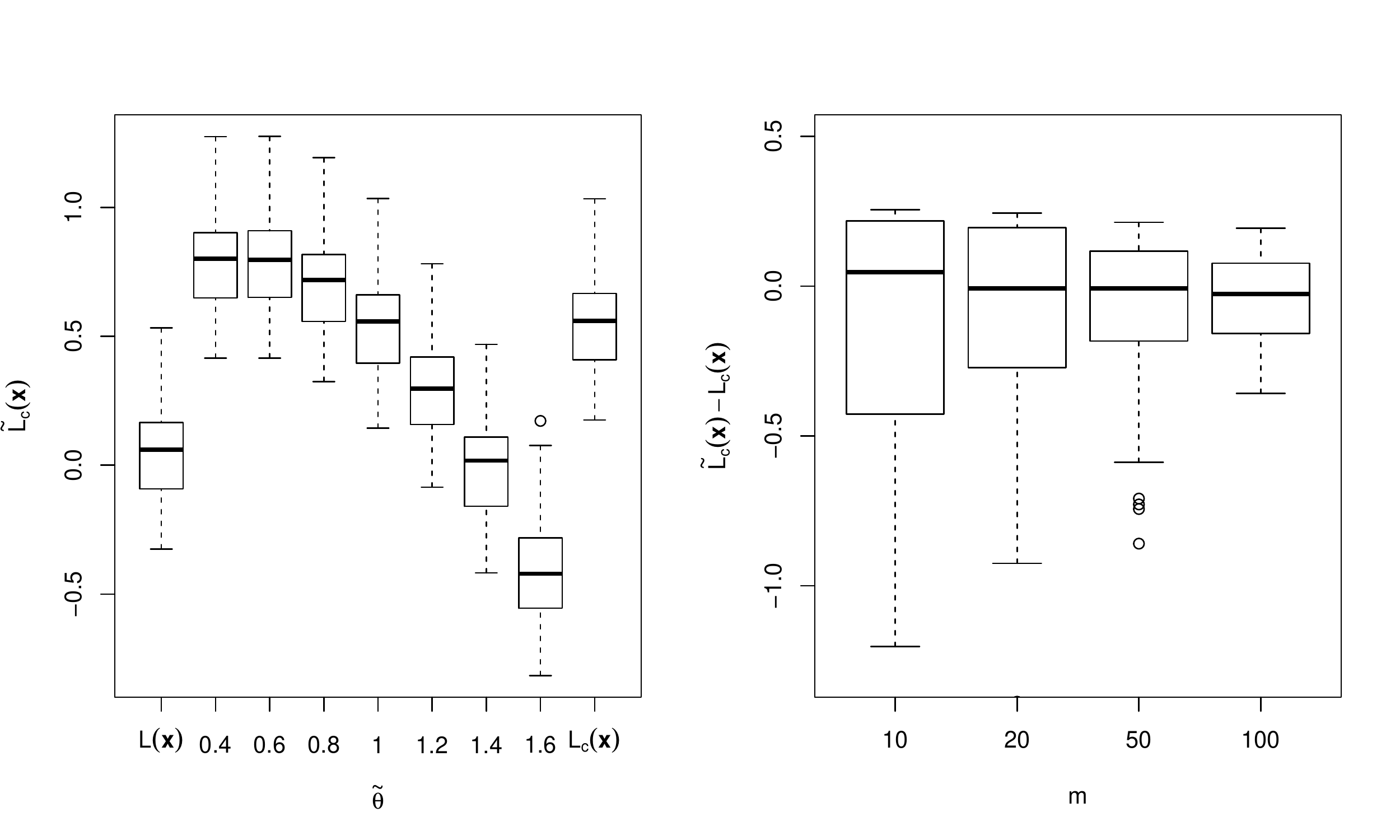}
\caption{ \small
Left panel: Boxplots of the corrected lower confidence limit, $\tilde{L}_c({\bf x})$, when holding $\tilde{\theta}$ fixed at various values $\tilde{\theta}=0.4, ..., 1.6$ (true value is $\tilde{\theta}=1)$. Leftmost and rightmost boxplots correspond to the biased ($L_{\bf x})$ and true unbiased ($L_c({\bf x})$) lower limits respectively.
Right panel: Boxplots of the correction error $\tilde{L}_c({\bf x})-L({\bf x})$ as a function of observed data sample size $m$.
All boxplots are based on 100 replicate analyses.
}
\label{newfigure2}
\end{center}
\end{figure}

\noindent Figure~\ref{newfigure2} shows the results of the correction procedure for the lower limit of a 95\% confidence interval based on 100 replicate analyses. Each analysis uses samples of size $m$ with $\theta=1$, so that ${x}_1,\ldots,{x}_m\sim N(0,1)$, and $n=2000$  replicated samples ${\bf y}_1,\ldots,{\bf y}_n$ with elements drawn from
$N(0,\tilde{\theta})$.
Figure~\ref{newfigure2} (left panel) illustrates the corrected lower confidence limit, $\tilde{L}_c({\bf x})$, based on a sample of size $m=20$, for a range of fixed values of $\tilde{\theta}$. The extreme left and right boxplots correspond to the raw biased ($L({\bf x})$) and true unbiased ($L_c({\bf x})$) limits respectively. Clearly, as $\tilde{\theta}$ changes, then so does the location of the adjusted limits. This occurs as, in contrast with the above example, the distributions of $\tilde{\theta}-L({\bf y})$ and $\tilde{\theta}-U({\bf y})$ clearly do change with $\tilde{\theta}$. When $\tilde{\theta}=\theta=1$, then the correction procedure produces the correct adjusted limits, as indicated by the rightmost boxplot. Hence, it is necessary to use the right value for $\tilde{\theta}$ when making the correction.
\\

\noindent Assumption 2 requires that $\tilde{\theta}$ is a consistent estimator of $\theta$. Hence we can be sure that $\tilde{\theta}\rightarrow\theta$ as $m\rightarrow\infty$, and as a result that the distribution of $\tilde{\theta}-L({\bf y})$ approaches that of $\theta-L({\bf x})$, so that  our correction procedure will perform correctly for large enough $m$. In practice, the required value of $m$ can be moderate.
\noindent Figure~\ref{newfigure2} (right panel) shows how the correction error, $\tilde{L}_c({\bf x})-L_c({\bf x})$, varies as a function of $m$. Clearly, the median error is close to zero even for small sample sizes. However, there is some asymmetry for small $m$, which is also visible in the left panel (e.g. compare the differences in the bias in the boxplots with $\tilde{\theta}=0.4$ and $\tilde{\theta}=1.6$), although this is eliminated as $m$ increases.
\\

\begin{table}[tbh]
\begin{center}
\begin{tabular}{|cccccc|cccccc|}
\hline
& \multicolumn{5}{c|}{Example 1: $\mu$} && \multicolumn{5}{c|}{Example 2: $\sigma^2$}\\
$\epsilon$&P& CP &B & CB & DB &$\epsilon$&P& CP &B & CB & DB\\
\hline
$0.00$ &0.956 &0.951&--&--&--&$0.00$& 0.954&0.954&--&--&-- \\
$ 6.67$&1.000 &0.954& --&--&--  &$0.20$& 0.939&0.952& --&--&--   \\
--& --&--&0.946&0.949&0.952&--&-- &--& 0.918 &0.956&0.964  \\
\hline
Time (s) &&0.09\phantom{1} & 0.21\phantom{0} & 4.53& 6.70\phantom{1}&&&0.17\phantom{0}&0.27\phantom{0}&7.31\phantom{0}&8.16\phantom{0}\\
\hline
\end{tabular}
\caption{\small Empirical coverage probabilities for 95\% confidence intervals for the parameters in  Examples 1 and 2, for various methods, and for differing values of error, $\epsilon$.
Columns denote coverage of (P) the pivot-based intervals, (CP) our correction of the pivot intervals, (B) parametric bootstrap intervals, (CB) our correction of the parametric  bootstrap intervals, and (DB) the double bootstrap intervals.
Coverage probabilities are based on $1,000$ replicate analyses under each method, and $n=2,000$ generated datasets ${\bf y}_1,\ldots,{\bf y}_n$ for the corrected pivotal (CP) and corrected bootstrap (CB).
The bootstrap results used 2,000 bootstrap samples (B), 99 bootstrap samples for the corrected bootstrap (CB), and 2000$\times$44 samples for the double bootstrap (DB).
 Time (in seconds) indicates the time needed to produce one adjusted replicate interval.
}
\label{tab:tablecov}
\end{center}
\end{table}

\noindent Finally, Table \ref{tab:tablecov} compares the empirical coverage probabilities for 95\% confidence intervals for both $\mu$ and $\sigma^2$ in Examples 1 and 2, using our correction procedure and the parametric bootstrap (e.g. \citeNP{davison1997bootstrap}).
\\

\noindent
Where there is no error in the construction of the pivot-based confidence intervals (column P) i.e. for $\epsilon=0$, the corrected intervals  (column CP) retain the same correct coverage properties as before the correction. However, we note that as the corrected intervals $(\tilde{L}_c({\bf x}),\tilde{U}_c({\bf x}))$ are estimated by Monte Carlo, for finite numbers of generated datasets, ${\bf y}_1,\ldots,{\bf y}_n$, there will be some non-zero adjustment of each individual confidence interval, even when no systematic error is present.  In Table \ref{tab:tablecov}, $n=2,000$ datasets were used for each corrected interval. However, in spite of this random adjustment, the correct coverage is retained over multiple replicates. This point is discussed further in Section 4.
The second row of Table \ref{tab:tablecov}, shows the same information as the first row, except with a non-zero error, $\epsilon=6.67$ (for $\mu$) and $\epsilon=0.20$ (for $\sigma^2$). Clearly the adjusted intervals have the correct nominal coverage.
\\

\noindent
The third row in Table \ref{tab:tablecov} illustrates the empirical coverage probabilites of 95\% confidence intervals based on using the parametric  bootstrap (B),  our correction of the parametric bootstrap (CB) and the double-bootstrap (DB). Each bootstrap (B) interval was based on 2,000 bootstrap samples, 2,000$\times$44 samples for the double bootstrap (DB) (following \citeNP{mccullough1998implementing,booth1994monte}),  99 samples for our correction of the bootstrap (CB), and $n=2000$ datasets, ${\bf y}_1,\ldots,{\bf y}_n$, for our correction procedure. These numbers were chosen to provide broadly comparable algorithmic overheads for each method. The bootstrap calculations were implemented using the R package {\tt boot}, and the double bootstrap confidence intervals were computed as in \citeN{mccullough1998implementing}.
\\

\noindent
The parametric bootstrap has previously been observed to have lower than nominal coverage (e.g. \citeNP{schenker85,buckland84}). In Table \ref{tab:tablecov} this is particularly apparent for $\sigma^2$. Both our correction and the double bootstrap produce improved coverage in each case, although the double bootstrap requires different specification (i.e. the number of bootstrap replicates at each of two levels) than our approach, which alternatively requires the number of auxiliary datasets, $n$, and a consistent estimator of $\theta$.
While it is  difficult to provide similar algorithmic specifications to permit speed comparisons between the double bootstrap and our correction procedure, in that they possess different algorithm structures, the recorded times for each method were broadly similar (Table \ref{tab:tablecov}, bottom row), with our procedure slightly faster in both current examples. However, the double bootstrap algorithm has a number of optimisations available (e.g. \citeNP{nankervis2005computational}), whereas our procedure was implemented with unoptimised code. Broadly, the two approaches are comparable in the present analyses.

\section{Real examples}%
\label{examples}

We now consider interval estimation in two real, complex modelling situations.
The first is an application of composite likelihood techniques  in the modelling of spatial extremes. With composite likelihoods, deriving unbiased confidence intervals can require a large amount of algebra, whereas biased intervals that are typically too narrow are easily computable.
The second is an application of approximate Bayesian computation (ABC) methods in the modelling of a time series of $g$-and-$k$ distributed observations. In most practical settings the mechanism behind the model fitting process within the ABC framework typically gives  posterior credible intervals that are too large.
\\

\noindent
In both analyses, the parameter $\theta$ is a vector of $d>1$ dimensions. However, our coverage correction procedure is a univariate method as it is based on quantiles. As such, after obtaining the consistent vector estimator $\tilde{\theta}$,
we correct the coverage probabilities of
each element of the parameter vector in turn, and obtain confidence intervals that achieve the correct nominal  marginal coverage probabilities.

\subsection{Spatial extremes via composite likelihoods}%

\noindent In the context of analysing spatial extremes, \shortciteN{padoan2010likelihood} developed a pairwise composite likelihood model, for inference using max-stable stationary processes. Specifically, for $m$ annual maximum daily rainfall observations, at each of $K$ spatial locations, the pairwise composite likelihood was specified as
\[
	\ell_C(\theta|{\bf x}) = \sum_{i<j} w_{ij}\log f({\bf x}_i,{\bf x}_j|\theta)
\]
where $\bf{x}=(\bf{x}_1,\ldots,\bf{x}_K)$ and ${\bf x}_i=({\bf x}_{i1},\ldots,{\bf x}_{im})$,
$f({\bf x}_i,{\bf x}_j|\theta)$ is a known bivariate density function with parameter vector $\theta$ evaluated at spatial locations $i$ and $j$, and
$w_{ij}>0$ are weights such that $\sum_{i,j}w_{ij}=1$.
Under the usual regularity conditions, the maximum composite likelihood estimator, $\hat{\theta}$, can provide asymptotically unbiased and normally distributed parameter estimates when standard likelihood estimators are unavailable (e.g. \shortciteNP{varin2011overview}).
\\

\noindent Specifically, we have (e.g.  \citeNP{huber1967behavior}) that $\hat{\theta} \sim N(\theta,\tilde{I}^{-1}(\hat{\theta}))$, with
\begin{equation}
\label{CMLE}
	{I}(\theta)=H(\theta)J(\theta)^{-1}H(\theta),
\end{equation}
where $H(\theta)$ and $J(\theta)$ are respectively the expected information matrix and the covariance matrix of the score vector. In the ordinary maximum likelihood setting,  $H(\theta)=J(\theta)$.
In the max-stable process framework, \shortciteN{padoan2010likelihood} provided an analytic expression for $J(\theta)$ for a particular (Gaussian) spatial dependence model. Combined with the standard numerical estimates of $H(\theta)$, this allowed for the construction of standard confidence intervals for $\theta$.
However, for composite likelihood techniques in general, obtaining analytic expressions or numerical estimates of $J(\theta)$ can be challenging, whereas estimates of $H(\theta)$ are readily available.
In this example, we demonstrate how our proposed method can be employed to correct the too narrow confidence intervals that result from using
${I}(\theta)=H(\theta)$.
We then compare our results with those derived from the known maximum composite likelihood information matrix (\ref{CMLE}).
\\

\noindent We considered four spatial models for stationary max-stable processes that describe different degrees of extremal dependence, with parameter inference based on $m=100$ observations at each of $K=50$ randomly generated spatial locations. Each model expresses the degree of extremal dependence via the covariance matrix
\[
	\Sigma=\left[
	\begin{array}{ll}
	\sigma^2_1 &\sigma_{12}\\
	\sigma_{12} &\sigma^2_2
	\end{array}\right],
\]
where
the values for each parameter for each model $M_1,\ldots,M_4$ are given in Table~\ref{table:tableSigma}.
Model $M_4$ has  an additional non-stationary spatial component that is modelled by the extra marginal parameters $\mu,\lambda$ and $\xi$ (corresponding to location, scale and shape parameters) through the response surface
\begin{eqnarray*}
\mu&=&\alpha_0+\alpha_1*lat+\alpha_2*lon\\
\lambda&=&\beta_0+\beta_1*lon\\
\xi&=&\gamma_0,
\end{eqnarray*}
where $lat$ and $lon$ denote latitude and longitude coordinates.
\\

\begin{table}[ht]
\centering
\begin{tabular}{|cccc|}
\hline
Model & $\sigma_{1}^2$ & $\sigma_{12}$ & $\sigma_{2}^2$\\
\hline
$M_1, M_4$ & 9/8&0 &9/8\\
$M_2$ & 2000&1200 &2800\\
$M_3$ & 25&35 &14\\
\hline
\end{tabular}
\caption{\small Covariance matrix configurations for models $M_1,\ldots,M_4$ for the extremal spatial dependence analysis.}
\label{table:tableSigma}
\end{table}
\noindent

\noindent
Tables~\ref{tab:tablesigmaestimates} and \ref{tab:tablesurface} summarise the empirical coverage probabilities for nominal $95\%$ confidence intervals for models $M_1,\ldots,M_4$ based on 500 replicate analyses.
Columns $C_1$ provide the interval coverage using the
standard Hessian matrix $I(\theta)=H(\theta)$, and columns $C_2$ provide the same using the composite likelihood information matrix  (\ref{CMLE}) following \shortciteN{padoan2010likelihood}.
Columns $C_3$ correspond to our correction procedure when applied to the intervals in column $C_1$ using the standard Hessian matrix. For the correction procedure we used $\tilde{\theta}=\hat{\theta}$, the maximum composite likelihood estimate, and $n=500$ simulated datasets to perform the adjustment.
\\

\noindent From Table \ref{tab:tablesigmaestimates}, clearly confidence interval coverage based on the standard Hessian matrix ($C_1$) is too low. The coverage using the sandwich information matrix ($C_2$) is very good, with all the reported values close to 0.95. The coverage values obtained using our adjustment procedure, which is based on the intervals in column $C_1$, are also very close to $0.95$, and mostly closer than with the sandwich information matrix. Similar results are obtained for model $M_4$ in Table \ref{tab:tablesurface}. Taken together, these results indicate that our adjustment procedure can successfully modify the upper and lower limits of a confidence interval to achieve comparable results to established methods in complex settings. However, it does not make use of the algebraic representation of $J(\theta)$ in this case, and so is more easily extended to alternative models (e.g. where $J(\theta)$ is not available), albeit at a moderate computational cost.

\begin{table}[tbh]
 \centering
\begin{tabular}{|c|ccc|ccc|ccc|}
\hline
 & \multicolumn{3}{c|}{$M_1$}& \multicolumn{3}{c|}{$M_2$} & \multicolumn{3}{c|}{$M_3$}\\
&$C_1$ &$C_2$ &$C_3$ &$C_1$ &$C_2$ &$C_3$  &$C_1$ &$C_2$ &$C_3$\\
\hline
$\sigma^2_{1}$&0.428&0.960 &0.960 &0.098 &0.947 &0.950&0.092 &0.939 & 0.944\\
$\sigma_{12}$ &0.518& 0.940  &0.960&0.122 &0.955 & 0.956&0.154 &0.921 &0.960\\
$\sigma^2_{2}$&0.468 &0.930 &0.936&0.092 & 0.955 &0.938&0.102 &0.940 &0.952\\
\hline
\end{tabular}
\caption{\small Empirical coverage probabilities for 95\% confidence intervals of the parameters of models $M_1,M_2$, and $M_3$ based on 500 replicate analyses. Columns indicate interval confidence estimation methods using: ($C_1$) the standard Hessian matrix $I(\theta)=H(\theta)$; ($C_2$) the sandwich information matrix $I(\theta)=H(\theta)J^{-1}(\theta)H(\theta)$; and ($C_3$) the standard Hessian matrix $I(\theta)=H(\theta)$ followed by our correction procedure.}
\label{tab:tablesigmaestimates}
\end{table}

\begin{table}[tbh]
\centering
\begin{tabular}{|cccccccccc|}
\hline
&$\sigma_{1}^2$&$\sigma_{12}$&$\sigma^2_{2}$&$\alpha_0$&$\alpha_1$ &$\alpha_2$&$\beta_0$&$\beta_1$&$\gamma_0$\\
\hline
$C_1$ &0.372 &0.544 &0.388 &0.114 &0.122 &0.098 &0.108 &0.140 &0.104\\
$C_2$ &0.930 &0.925 &0.945 &0.935 &0.945 &0.945 &0.935 &0.940 &0.910\\
$C_3$ &0.924 &0.924 &0.958 &0.950 &0.940 &0.944 &0.944 &0.952 &0.952\\
\hline
\end{tabular}
\caption{\small Empirical coverage probabilities for 95\% confidence intervals of the parameters of model $M_4$ based on 500 replicate analyses.  Columns indicate interval confidence estimation methods using: ($C_1$) the standard Hessian matrix $I(\theta)=H(\theta)$; ($C_2$) the sandwich information matrix $I(\theta)=H(\theta)J^{-1}(\theta)H(\theta)$; and ($C_3$) the standard Hessian matrix $I(\theta)=H(\theta)$ followed by our correction procedure.}
\label{tab:tablesurface}
\end{table}

\subsection{Exchange rate analysis using approximate Bayesian computation}%

Approximate Bayesian computation (ABC) describes a family of methods of approximating a posterior distribution when the likelihood function is computationally intractable, but where sampling from the likelihood is possible (e.g.  \shortciteNP{beaumont2002approximate,sisson+f11}).
These methods can be thought of as constructing a conditional density estimate of the posterior \cite{blum10}, where the scale parameter, $h>0$, of the kernel density function controls both the level of accuracy of the approximation, and the computation required to construct it. Lower $h$ results in more accurate posterior approximations, but in return requires considerably more computation. As such, moderate values of the scale parameter are often used in practice. Accordingly, this typically results in oversmoothed estimates of the posterior, and in turn, credible intervals that are too wide.
\\

\noindent We consider an analysis of daily exchange rate log returns of the British pound to the Australian dollar between 2005 and 2007.
\shortciteN{drovandi2011likelihood} developed an MA(1) type model for these data where the individual log returns were modelled by a $g$-and-$k$ distribution \cite{rayner2002numerical}. The $g$-and-$k$ distribution is typically defined through it's quantile function
\begin{eqnarray}
\label{eqn:gandk}
 Q(z(p);\theta)=a+b\left(1+c\frac{1-\exp(-gz(p))}{1+\exp(-gz(p))}\right)(1+z(p)^2)^kz(p),
\end{eqnarray}
where $\theta=(a, b,g,k)$ are parameters controlling location, scale, skewness and kurtosis, and $z(p)$ is the $p$-quantile of a standard normal distribution. The parameter $c=0.8$ is typically fixed.
We used the sequential Monte Carlo-based ABC algorithm in \shortciteN{drovandi2011likelihood}, based on 2,000 particles, to fit the MA(1) model.
The data-generation process, used in both ABC and our correction procedure,  consists of drawing dependent quantiles $z_i=(\eta_i+\alpha\eta_{i-1})/\sqrt{1+\alpha^2}$ for $i=1,\ldots n$, where $\eta_i\sim N(0,1)$ for $i=0,\ldots,n$, and then substituting $z(p)=z_i$ in (\ref{eqn:gandk}).
\\

\noindent Table~\ref{tab:tableabcre} shows the estimated 95\% central credible intervals, and their widths, for each model parameter based on the ABC kernel scale parameter $h=0.016$ (following \citeNP{drovandi2011likelihood}) and also the lower value of $h=0.009$.
Also shown are the intervals obtained after performing a local-linear, ridge regression-adjustment \shortcite{blum+nps12,beaumont2002approximate} on the posterior obtained with $h=0.016$. The regression-adjustment is a standard ABC technique for improving the precision of an ABC posterior approximation, which aims to estimate the posterior at $h=0$ based on an assumed regression model.
\\

\noindent Clearly the parameter credible intervals obtained with $h=0.009$ are narrower than those obtained with $h=0.016$, indicating that the larger intervals indeed have greater than 95\% coverage. The regression-adjusted intervals generally have widths somewhere between the intervals constructed  with $h=0.016$ and $h=0.009$. The suggestion from Table~\ref{tab:tableabcre} is that even narrower (i.e. more accurate) credible intervals may result if it were possible to reduce $h$ further.
\\

\noindent Table~\ref{tab:tableabcC} shows the corrected 95\% central credible interval estimates, obtained from the ABC posterior approximations with kernel scale parameter $h=0.016, 0.02$ and $0.03$. The correction was based on $n=500$ simulated datasets and using the posterior mean as the estimate $\tilde{\theta}$ of $\theta$.
The results of the correction across the three kernel scale parameter values are similar, suggesting potential computational savings in the ABC posterior simulation stage, as one may perform the analysis with larger values of $h$.
All parameters achieve equivalent or improved precision compared to the most precise ABC posterior estimate obtained with $h=0.009$.
\\

\noindent  While for a standard Bayesian analysis, the posterior mean is a consistent estimator of $\theta$, this may not be true in the case of the ABC approximate posterior for $h>0$, as the location and shape of the ABC posterior can change with $h$. However, the posterior mean is a consistent estimator for $\theta$ for $h=0$. As such, some care may be needed when specifying $\tilde{\theta}$ as the posterior mean
in the ABC setting. In the current analysis, a preliminary investigation suggested that estimates of the posterior mean stabilised below $h=0.03$, which suggest that the posterior mean is approximately consistent for $h<0.03$. 
While this determination is slightly ad-hoc, it is practically viable, and an intuitively sensible way of determining whether the computed posterior mean is a consistent estimator of $\theta$. As such, we are confident that the posterior mean produces an effectively consistent estimate $\tilde{\theta}$ of $\theta$ in this case.

\begin{table}[tbh]
\begin{center}
\begin{tabular}{|c|cc|cc|cc|}
\hline
 & $h=0.016$ &Width & $h=0.009$ &Width & Reg. Adj. ($h=0.016$) &Width\\
\hline
$a$&(-0.0006, 0.0002) &0.0008 & (-0.0004,  0.0001) &0.0005 &(-0.0006, 0.0002) &0.0008\\
$b$&(\phantom{-}0.0018, 0.0028)&{0.0010} &(\phantom{-}0.0019, 0.0026) &{0.0007}&(\phantom{-}0.0018,  0.0027) &{0.0009}\\
$g$&(-0.0267, 0.2573)&{0.2840}&(-0.0044, 0.2138) &{0.2182}&(-0.0286, 0.2505) &{0.2791}\\
$k$&(\phantom{-}0.2024, 0.5061) &{0.3037 }&(\phantom{-}0.2607, 0.5322) &{0.2715} &(\phantom{-}0.2148,   0.5092)&{0.2944}\\
$\alpha$&(\phantom{-}0.1413, 0.2713)&{0.1300}&(\phantom{-}0.1491, 0.2771)&{0.1280}&(\phantom{-}0.1489, 0.2742) &{0.1253}\\
\hline
\end{tabular}
\caption{95\% central credibile intervals and corresponding interval widths from the  $g$-and-$k$ distribution MA(1) model. Results obtained using ABC posterior approximation with kernel scale parameter $h=0.016$ and $h= 0.009$, and following a ridge regression-adjustment based on an ABC posterior approximation with $h=0.016$.}
\label{tab:tableabcre}
\end{center}
\end{table}

\begin{table}[tbh]
\begin{center}
\begin{tabular}{|c|cc|cc|cc|}
\hline
 & $h=0.016$ &Width & $h=0.02$ &Width  & $h=0.03$ &Width\\
\hline
$a$&(-0.0003, 0.0000) &{0.0003}&(-0.0003, 0.0000) &{0.0003}&(-0.0004, -0.0001)&{0.0005} \\
$b$&(\phantom{-}0.0020, 0.0024)&{0.0004}&(\phantom{-}0.0021, 0.0025) &{0.0004}&(\phantom{-}0.0021, \phantom{-}0.0024) &{0.0003}\\
$g$&(\phantom{-}0.0303, 0.2156)&{0.1853}&(\phantom{-}0.0173, 0.1818)&{0.1645 }&(\phantom{-}0.0204, \phantom{-}0.1957) &{0.1753}\\
$k$&(\phantom{-}0.2769, 0.4099)&{0.1330 }&(\phantom{-}0.2909, 0.4235) &{0.1326 }&(\phantom{-}0.2768, \phantom{-}0.4129)&{0.1362}\\
$\alpha$&(\phantom{-}0.1430, 0.2659)&{0.1229}&(\phantom{-}0.1335, 0.2708)&{0.1373}&(\phantom{-}0.1513, \phantom{-}0.2714) &{0.1201}\\
\hline
\end{tabular}
\caption{Adjusted 95\% central credibility intervals and corresponding interval widths  from the $g$-and-$k$ distribution MA(1) model. Adjusted intervals based on correcting ABC posterior approximations with kernel scale parameter $h=0.016, 0.02$ and  $0.03$.}
\label{tab:tableabcC}
\end{center}
\end{table}

\section{Discussion}%
\label{secConc}

In this article we have introduced a method of adjusting confidence interval estimates to have a correct nominal coverage probability. This method was developed in the frequentist framework, but may be equally applied to ensure that Bayesian credible intervals possess the (frequentist) coverage property.
Our approach is general and makes minimal assumptions: namely that it is possible to generate data under the same procedure (model) that produced the observed data, and that a consistent estimator is available for the parameter of interest. The correction is asymptotically unbiased, although it can work well for moderate sample sizes ($m$), and there is a central limit theorem for the corrected interval limits in terms of the number ($n$) of auxiliary samples used to implement the correction.
\\

\noindent
As the correction is estimated by Monte Carlo, when there is no bias present, so that $L({\bf x})=L_c({\bf x})$,
for finite numbers of replicate datasets ${\bf y}_1,\ldots,{\bf y}_n$, 
finite sample estimates of  $\xi_{\alpha/2}$  may be non-zero. This will result in small, non-zero adjustments to intervals that already have the correct nominal coverage. In practice, for moderate $n$, this is likely to have negligible effect (e.g. see the results in Table \ref{tab:tablecov}). However, in this and other settings where there is low bias, the central limit theorem of Corollary 1
describes the precision of the finite sample adjustment as a function of $n$, thereby providing a guide as to when the Monte Carlo variability of $\tilde{L}_c({\bf x})$ will be an improvement over the bias of $L({\bf x})$. \\

\noindent
As constructed in Theorems 2.1 and 2.2, our proposed correction is for univariate parameters, $\theta$, as it is based on quantiles.
For multivariate $\theta$, from the perspective of adjusting any given margin,
the impact of the remaining (nuisance) parameters is controlled through the estimate $\tilde{\theta}$ of $\theta$. 
Asymptotically, the consistency of $\tilde{\theta}$ means that $\tilde{\theta}\rightarrow\theta$ as the sample size $m\rightarrow\infty$, from which Theorems 2.1 and 2.2 will then hold for the margin of interest. However, sub-asymptotically this is not the case, and the performance of the adjustment of any margin will depend on the quality of the estimate of $\theta$ (this is also true in the univariate setting). The results of our analyses in Sections \ref{sec:simple} and \ref{examples} suggest that the procedure can work well, even for moderate $m$.
\\

\noindent In practice, in the examples that we have considered, we have found that our method can produce confidence intervals which perform comparably to existing gold standard approaches -- though with greater scope for extension to more complicated models -- and provide a reliable method of adjusting approximately obtained credible intervals in challenging settings.
\\

\noindent One potential criticism of our approach is that it requires the construction of a large number ($n$) of confidence or credible  intervals in order to correct one interval. In the case where constructing a single interval is computationally expensive, implementing the correction procedure in full can result in a large amount of computation. This was the case in our exchange rate data analysis using ABC methods,  where using an alternative ABC algorithm such as regression-adjustment (based on a single large number of model simulations) would have been more efficient.
However, regression-adjustment can itself perform poorly if the assumed regression model is incorrect, while as our correction procedure makes minimal assumptions, we may still have good confidence  in the resulting adjusted intervals it provides.

\section*{Acknowledgements}

We are grateful to Chris Drovandi and Tony Pettitt for kindly allowing us to use their Matlab codes for the ABC analysis. Financial support for this research was provided by the Australian Research Council Discovery Project scheme (DP1092805) and the University of New South Wales.

\section*{Appendix: Proofs}%

\subsection*{Proof of Theorem \ref{thm1}}%

Writing $L_c({\bf x}) = L( {\bf x}) + \delta$, then
\begin{eqnarray*}
P ( \theta < L_c({\bf x}) ) &=& P (\theta < L( {\bf x}) + \delta ) \\
&=&P ( \theta- L( {\bf x}) < \delta ).
\end{eqnarray*}
Hence, by definition, $P ( \theta < L_c({\bf x}) ) = \alpha/2$ if
$\delta=\xi_{\alpha/2}$, where $\xi _{\alpha/2}$ is the $\alpha/2$-th quantile of the distribution of $\theta-L({\bf x})$.

\subsection*{Proof of Theorem \ref{thm2}}%

Let  $G_{ \{ \theta-L({\bf x}) \} }$
be the distribution function  of $\theta-L({\bf x})$, which has positive first derivatives so that $G_{ \{ \theta-L({\bf x}) \} }^{'}(\nu)=\frac{\partial}{\partial z} G_{ \{ \theta-L({\bf x}) \} }(\nu)>0$ for all $\nu\in{\mathbb R}$.
Also let $\hat{G}_{ \{ \tilde{\theta}-L({\bf y}) \} }$ be the empirical distribution of $\tilde{\theta}-L({\bf y})$ based on the samples $\tilde{\theta}-L_i({\bf y}_i)$, $i=1,\ldots, n$.
From Theorem \ref{thm1} we have that
$L_c({\bf x})=L({\bf x})+\xi_{\alpha/2}$ for some $\alpha\in(0,1)$, where  $G_{ \{ \theta-L({\bf x}) \} }(\xi_{\alpha/2})=\alpha/2$.
Let $\hat{\xi}_{\alpha/2} = \hat{G}^{-1}_{ \{ \tilde{\theta}-L({\bf y})\}} (\alpha/2)$ be the empirical estimate of $\xi_{\alpha/2}$.
\\

\noindent If we define  $\tilde{L}_c({\bf x})=L({\bf x})+\hat{\xi}_{\alpha/2}$, then for any $\omega\in\mathbb{R}$, we have
\begin{eqnarray*}
\mbox{Pr}\Big(\sqrt{n}(\tilde{L}_c({\bf x})-L_c({\bf x}))\leq \omega)&&=\mbox{Pr}\Big(\sqrt{n}(\hat{\xi}_{\alpha/2}-\xi_{\alpha/2})\leq \omega)\\
&&=\mbox{Pr}(\hat{\xi}_{\alpha/2}\leq \xi_{\alpha/2}+\omega/\sqrt{n}\Big)\\
&&=\mbox{Pr}\Big(G_{\{ \theta-L({\bf x}) \}}(\hat{\xi}_{\alpha/2})\leq G_{\{ \theta-L({\bf x}) \}}(\xi_{\alpha/2}+\omega/\sqrt{n})\Big)\\
&&=\mbox{Pr}\Big(G_{\{ \theta-L({\bf x}) \}}(\hat{\xi}_{\alpha/2})\leq \alpha/2+[\omega G_{\{ \theta-L({\bf x}) \}}^{'}(\xi_{\alpha/2})+o(1)]/\sqrt{n}\Big)
\end{eqnarray*}
where the last equality follows from a first order Taylor expansion of $G$ at $\xi_{\alpha/2}$.
\\

\noindent If $Y$ represents the number of times that $G(\hat{\xi}_{\alpha/2})$ is smaller than $\zeta=\alpha/2+[\omega G_{\{ \theta-L({\bf x}) \}}^{'}(\xi_{\alpha/2})+o(1)]/\sqrt{n}$, then since $G(\xi)  \sim U(0,1)$, we have $Y\sim \mbox{Binomial}(n,\zeta)$. Hence
\begin{eqnarray}\label{eqnBIN}
\frac{Y-n\zeta}{\sqrt[]{n\zeta(1-\zeta)}}\rightarrow N(0,1)
\end{eqnarray}
in distribution as $n\rightarrow\infty$ \cite{van2000asymptotic}.
\\

\noindent Let $r_n$ be the integer rank of the $\alpha/2$-th quantile from a data set $X=\{X_1,\ldots,X_n\}$ of length $n$, such that $\hat{\xi}_{\alpha/2}=X_{(r_n)}$. If we assume that $\frac{n\alpha/2-r_n}{\sqrt[]{n}}\rightarrow 0$ as $n\rightarrow\infty$, then from (\ref{eqnBIN}) we have
\begin{eqnarray}
\mbox{Pr}\Big(\sqrt{n}(\hat{\xi}_{\alpha/2}-\xi_{\alpha/2})\leq \omega\Big)&&=\mbox{Pr}\Big(G_{\{ \theta-L({\bf x}) \}}(\hat{\xi}_{\alpha/2})\leq \zeta\Big)=\mbox{Pr}\Big(Y\geq r_n\Big)\nonumber\\
&&=\mbox{Pr}\left(\frac{Y-n\zeta}{\sqrt[]{n\zeta(1-\zeta)}}\geq \frac{r_n-n\zeta}{\sqrt[]{n\zeta(1-\zeta)}}\right)\nonumber\\
&&=\Phi\left(\frac{n\alpha/2+\sqrt{n}\omega G_{\{ \theta-L({\bf x}) \}}^{'}(\xi_{\alpha/2})-r_n}{\sqrt[]{n\zeta(1-\zeta)}}\right)+o_p(1)\nonumber\\
&&=\Phi\left(\frac{n (\alpha/2)-r_n}{\sqrt[]{n(\alpha/2)(1-\alpha/2)}}+\frac{\omega G_{\{ \theta-L({\bf x}) \}}^{'}(\xi_{\alpha/2})}{\sqrt[]{(\alpha/2)(1-\alpha/2)}}\right)+o_p(1)\nonumber\\
&&=\Phi\left(\frac{\omega G_{\{ \theta-L({\bf x}) \}}^{'}(\xi_{\alpha/2})}{\sqrt[]{(\alpha/2)(1-\alpha/2)}}\right) +o_p(1).\label{eqnCLT}
\end{eqnarray}
It then follows that the consistency of the estimator $\tilde{L}_c$ can be established as
$$\underset{n\rightarrow \infty}{\lim} \mbox{Pr}(\sqrt{n}(\tilde{L}_c({\bf x})-L_c({\bf x})) \geq \epsilon) \leq \underset{n\rightarrow\infty}{\lim}\frac{E[\sqrt{n}(\tilde{L}_c({\bf x})-L_c({\bf x}))]}{\epsilon} = 0$$
by the Markov inequality \cite{ash2000probability}.

\newpage
\bibliographystyle{chicago}
\bibliography{biascorrection2}
\end{document}